\pdfoutput=1 
\documentclass[dvipdfm]{PoS}

\usepackage{graphicx}
\usepackage{amssymb,amsfonts,amsmath}

\usepackage{fancyhdr}
\usepackage{setspace}
\usepackage{tikz}
\def\beq{\begin{equation}}
\def\eeq{\end{equation}}
\def\bea{\begin{eqnarray}}
\def\eea{\end{eqnarray}}

\def\beq{\begin{equation}}
\def\eeq{\end{equation}}
\def\bea{\begin{eqnarray}}
\def\eea{\end{eqnarray}}

\newcommand{\hp}{\hat{p}}
\newcommand{\hk}{\hat{k}}

\newcommand{\uphi}{{\Phi}}
\newcommand{\ups}{{\Psi_1}}
\newcommand{\upS}{{\Psi_2}}

\newcommand{\kakuk}[1]{\left[ #1 \right]}
\newcommand{\maruk}[1]{\left( #1 \right)}

\newcommand{\half}[0]{\frac{1}{2}}

\newcommand{\invs}[1]{\frac{1}{#1}}

\newcommand{\braket}[1]{\left< #1 \right>}

\newcommand{\ovphi}[0]{\overline{\varphi}}
\newcommand{\ovf}[0]{\overline{f}}
\newcommand{\ovpsi}[0]{\overline{\psi}}

\title{Exact Lattice Supersymmetry at the Quantum Level for
N=2 Wess-Zumino models in Lower Dimensions}

\ShortTitle{Ward-Takahashi Identity for Exact Lattice SUSY Wess-Zumino Models}

\author{Keisuke ~Asaka\\
Department of Physics, Hokkaido University\\
Sapporo, 060-0810 Japan\\
E-mail: \email{asaka@particle.sci.hokudai.ac.jp}}

\author{Alessandro~D'Adda\\
        INFN Sezione di Torino and
Dipartimento di Fisica Teorica,
Universit\`a di Torino\\
I-10125 Torino, Italy\\
        E-mail: \email{dadda@to.infn.it}}

\author{\speaker{ Noboru~Kawamoto}\\
        Department of Physics, Hokkaido University\\
Sapporo, 060-0810 Japan\\
        E-mail: \email{kawamoto@particle.sci.hokudai.ac.jp}}

\author{Yoshi ~Kondo\\
Department of Physics, Hokkaido University\\
Sapporo, 060-0810 Japan\\
E-mail: \email{proton@particle.sci.hokudai.ac.jp}}

\abstract{We have recently proposed a new lattice SUSY formulation
which has exact lattice supersymmetry for Wess-Zumino models in
one and two dimensions for all N=2 supercharges. This formulation
is non-local in the coordinate space but the difference operator
satisfies the Leibniz rule on the newly defined star product. Here
we show that this lattice supersymmetry is kept exact at the
quantum level by investigating Ward-Takahashi identities up to two
loop level.  }

\FullConference{The 30 International Symposium on Lattice Field Theory - Lattice 2012,\\
        June 24-29, 2012\\
        Cairns, Australia}

\begin{document}

\section{Introduction}

There are two major difficulties in constructing exact lattice
SUSY
formulation for all super charges:\\
1) The difference operator does not satisfy the Leibniz rule.\\
2) For massless lattice fermions  species doublers of
chiral fermions usually appear. \\
If we replace the differential operator by a difference operator
in the SUSY algebra, lattice SUSY is broken at the algebraic level
since the SUSY generators satisfy Leibniz rule while the
difference operator does not follow to the Leibniz rule. Secondly
if we put massless fermions on the lattice species doublers of the
chiral fermion  appear:  an unavoidable consequence of the NO-GO
theorem of  chiral fermions on the lattice. In supersymmetry the
number of boson degrees of freedom and that of fermions should be
the same, and thus this chiral fermion doublers break the balance
of degrees of freedom between the bosons and fermions. Thus
lattice supersymmetry will be broken with the naive version of
lattice fermion formulation. Even if we use the recently proposed
chiral fermion formulation satisfying Ginzberg-Wilson relation,
the treatment of fermions and bosons cannot be exactly the same
leading to a breaking of exact lattice supersymmetry. It has
recently been pointed out that the item 1) is in fact a NO-GO for
local lattice formulation of supersymmetry\cite{Kato:2008sp}.

With the aim of solving these difficulties we proposed the
formulation of ref. \cite{D'Adda:2010pg}\cite{D'Adda:2012}. For
the problem 1) we identify the momentum representation of a
symmetric lattice difference operator as a lattice momentum and
impose the conservation of the lattice momenta for products of
fields in the momentum representation. The importance of the
lattice momentum conservation was noticed by the very first paper
of lattice SUSY\cite{D-N}. In solving the problem 2) we identify
the species doublers as super partner particles in the same super
multiplet. To keep the balance for the equal treatment of fermions
and bosons we introduce the species doubler counter part for
bosons. We briefly explain the lattice SUSY formulation N=2
Wess-Zumino model in two dimensions, which has exact lattice
SUSY\cite{D'Adda:2012}. We explicitly show that the exact SUSY is
kept at the quantum level by explicitly examining the
Ward-Takahashi (WT) identities up to two loop level. One
dimensional formulation of Wess-Zumino model which has exact
lattice SUSY is given in \cite{D'Adda:2010pg}.

\section{D=N=2 Wess-Zumino action}

$N=2$ extended supersymmetry algebra in two dimensions is given by
\beq
\{Q_{\alpha i}, Q_{\beta j}\}~ = ~2\delta_{ij}
(\gamma^\mu)_{\alpha \beta} i\partial_\mu ,
\label{D=N=2-alg1}
\eeq
where we may use an explicit representation of Pauli matrices for
$\gamma^\mu=\{\sigma^3,\sigma^1\}$. By going to the light cone
directions this two dimensional $N=2$
algebra can be decomposed into the direct sum of two one dimensional $N=2$ algebra :
\beq
\{Q_\pm^{(i)},Q_\pm^{(j)}\} ~=~ 2\delta^{ij} i\partial_\pm, ~~~~~~
\{\hbox{others}\} =0,
\label{D=N=2-alg2}
\eeq
where
\beq
Q_\pm^{(j)}~=~\frac{Q_{1j}\pm i Q_{2j}}{\sqrt{2}}, ~~~~~~
\partial_\pm = \partial_1 \pm i\partial_2 ,
\label{def-scharge-N=D=2-1} \eeq Here we have introduced the
following light cone coordinates \beq x_\pm=x_1\pm ix_2,~~~~~~
\partial_\pm=\frac{\partial}{\partial x_\pm}.
\label{euclid-light-cone} \eeq We can equivalently express the
above algebra in a chiral form: \beq
\{Q_{\pm}^{(+)},Q_{\pm}^{(-)}\} ~=~i\partial_{\pm},
~~~\{\hbox{others}\}=0, \label{D=N=2-alg3} \eeq where \beq
Q_\pm^{(+)} = \frac{Q_\pm^{(1)} + iQ_\pm^{(2)}}{2}, ~~~
Q_\pm^{(-)} = \frac{Q_\pm^{(1)} - iQ_\pm^{(2)}}{2}.
\label{def-scharge-N=D=2-2} \eeq The corresponding momentum
counterpart of the algebra is given by: \beq
\{Q_{\pm}^{(+)},Q_{\pm}^{(-)}\} ~=~2\sin\frac{ap_{\pm}}{2}
~\equiv~\hat{p}_{\pm}. \label{D=N=2-mom-alg3} \eeq In two
dimensional N=2 SUSY algebra, we introduce four chiral fields
$\uphi_A \equiv \{\uphi,\ups,\upS,F\}$ and the corresponding
anti-chiral fields $\overline{\uphi}_A $. Each field $\uphi_A$ and
$\overline{\uphi}_A $ has 4 species doublers. We can impose chiral
and anti-chiral conditions which lead to the identification of the
original fields with the species doubler fields
\cite{D'Adda:2012}: \bea
\uphi_A(p_+,p_-) &=& \uphi_A(\frac{2\pi}{a}-p_+,p_-)=\uphi_A(p_+,\frac{2\pi}{a}-p_-)=\uphi_A(\frac{2\pi}{a}-p_+,\frac{2\pi}{a}-p_-), \nonumber \\
\overline{\uphi}_A(p_+,p_-) &=& -\overline{\uphi}_A(\frac{2\pi}{a}-p_+,p_-)
=-\overline{\uphi}_A(p_+,\frac{2\pi}{a}-p_-)=
\overline{\uphi}_A(\frac{2\pi}{a}-p_+,\frac{2\pi}{a}-p_-)
\label{chiral2}
\eea
with $\uphi_A \equiv \{\uphi,\ups,\upS,F\}$.

The kinetic term of the supersymmetric Wess-Zumino action can be written
in a $Q-$exact form of action as in the continuum:
\bea
S_K &=&\int_{-\frac{\pi}{a}}^{\frac{3\pi}{a}} dp_+ dp_- dq_+ dq_-  \delta(\hat{p}+\hat{q})  Q_+^{(-)}Q_-^{(-)}Q_+^{(+)}Q_-^{(+)} \{\bar{\uphi}(p)\uphi(q)\}
\nonumber \\
&=&\int_{-\frac{\pi}{a}}^{\frac{3\pi}{a}} dp_+ dp_- dq_+ dq_-  \delta(\hat{p}+
\hat{q})
 \left[-4\bar{\uphi}(p) \sin\frac{aq_+}{2}\sin\frac{aq_-}{2}\uphi(q) -\bar{F}(p) F(q)\right].\nonumber \\
 &&\left.+2\bar{\upS}(p)\sin\frac{aq_+}{2} \upS(q) +
2\bar{\ups}(p)\sin\frac{aq_-}{2}\ups(q)\right].
\label{DN2-WZ-action}
\eea
The invariance of the action $S_K$ under all the supersymmetry transformations
generated by $Q_{\pm}^{(\pm)}$ is assured by the algebra of
(\ref{D=N=2-mom-alg3}) whose component representation is given in
Tables 1 and 2 and by the momentum conservation
for the lattice momentum: $\hat{p}=2\sin \frac{ap}{2}$,
\beq
\delta(\hat{p}+\hat{q}) \equiv \prod_{i=\pm} \frac{1}{2}\left[ \delta(p_i+q_i)+\delta(p_i-q_i+\frac{2\pi}{a}) \right],~~~~~~~~(\textrm{mod}\frac{4\pi}{a}).
\label{lcons}
\eeq
\begin{table}% [c]
\hfil
$
\def\arraystretch{1.2}
\begin{array}{|l|c|c|c|c|} \hline
   &\displaystyle  Q_+^{(+)} & \displaystyle Q_+^{(-)}
&\displaystyle Q_-^{(+)} &\displaystyle Q_-^{(-)} \\ \hline
\hline
\uphi(p) & i\ups(p) & 0& i\upS(p) & 0 \\
\ups(p) & 0 & -2i\sin \frac{ap_+}{2} \uphi(p) & -F(p) & 0 \\
\upS(p) & F(p) &  0 & 0 & -2i\sin \frac{ap_-}{2}\uphi(p) \\
F(p) & 0 & 2\sin \frac{ap_+}{2} \upS(p) & 0
&-2\sin \frac{ap_-}{2} \ups(p)\\ \hline
\end{array}
$
\caption{Chiral $D=N=2$ supersymmetry transformation }
\end{table}

\begin{table}% [c]
\hfil
$
\def\arraystretch{1.2}
\begin{array}{|l|c|c|c|c|} \hline
   &\displaystyle  Q_+^{(+)} & \displaystyle Q_+^{(-)}
&\displaystyle Q_-^{(+)} &\displaystyle Q_-^{(-)} \\ \hline
\hline
\bar{\uphi}(p) & 0 & i\bar{\ups}(p) & 0  & i\bar{\upS}(p)  \\
\bar{\ups}(p) & -2i\sin \frac{ap_+}{2} \bar{\uphi}(p) & 0 & 0 & -\bar{F}(p)  \\
\bar{\upS}(p) & 0& \bar{F}(p) &  -2i\sin \frac{ap_-}{2}\bar{\uphi}(p)& 0 \\
\bar{F}(p) & 2\sin \frac{ap_+}{2} \bar{\upS}(p) & 0
&-2\sin \frac{ap_-}{2} \bar{\ups}(p) & 0 \\ \hline
\end{array}
$
\caption{anti-chiral $D=N=2$ supersymmetry transformation }
\end{table}

Interaction terms can be obtained by $Q$-exact form of the following action:
\bea
S_n &=&\int \prod_{j=1}^n  d^2p_j V_n(p)
Q_+^{(+)} Q_-^{(+)}\{ \uphi(p_1)\uphi(p_2) \cdots \uphi(p_n) \} + \textrm{h.c.}
\label{int1} \\
&=& \int \prod_{j=1}^n  d^2p_j V_n(p) n \left[i F(p_1)\prod_{j=2}^n \uphi(p_j) +(n-1) \upS(p_1)\ups(p_2) \prod_{j=3}^n \uphi(p_j) \right] +\textrm{h.c.},
\nonumber
\eea
where $V_n(p) $ is
\beq
V_n(p) = a^{2n}~ g_n~ G_n(p)~\delta^{(2)}\left(\sin\frac{ap_1}{2}+\sin\frac{ap_2}{2}+\cdots+\sin\frac{ap_n}{2}\right),
\label{sdelta}
\eeq
with $G_n(p)$ as appropriate momentum function which does not affect to the
lattice SUSY invariance.

We assume that all fields satisfy the (anti-) chiral conditions (\ref{chiral2}), so that in each variable the contribution of the integration in the intervals $(-\frac{\pi}{a},\frac{\pi}{a})$ and $(\frac{\pi}{a},\frac{3\pi}{a})$ coincide and we get
\bea
S_K &=& 4 \int_{-\frac{\pi}{a}}^{\frac{\pi}{a}} dp_+ dp_- dq_+ dq_-  \delta(p_+ +q_+)\delta(p_- +q_-)
 \left[-4\bar{\uphi}(p) \sin\frac{aq_+}{2}\sin\frac{aq_-}{2}\uphi(q) \right.\nonumber \\
 &&\left. -\bar{F}(p) F(q)+2\bar{\upS}(p)\sin\frac{aq_+}{2} \upS(q) +
2\bar{\ups}(p)\sin\frac{aq_-}{2}\ups(q)\right].
\label{DN2-WZ-action2}
\eea
The mass term in momentum representation is given as:
\beq
S_2 = m a^2  \int \prod_{j=1}^2 d^2p_j~ \delta(\hp_1+\hp_2) \left[ i F(p_1)\uphi(p_2)+ \upS(p_1)\ups(p_2) \right], \label{mass1}
\eeq
where the chiral conditions (\ref{chiral2}) are imposed.

 The dimensionless chiral fields can be rescaled with powers of the lattice constant $a$
 to match the canonical dimensions of the component fields:
\begin{align}
\uphi (p) &\to a^{-2} \varphi (p), & \Psi_i (p) & \to
a^{-\frac{3}{2}} \psi_i (p), & F (p) & \to a^{-1} f(p).
\end{align}
The anti-chiral fields are  similarly rescaled. It is also
necessary to rescale supercharges to recover correct canonical
dimension: $Q_i^{(j)} \to a^\half Q^{ (j)}_i$. The kinetic term in
momentum representation then reads:
\begin{align}
S_k &= \int^{\frac{\pi}{a}}_{-\frac{\pi}{a}} \frac{d\hp^2}{(2\pi)^2} \kakuk{ -\ovphi (-p) \hp_+ \hp_- \varphi (p) - \ovf (-p) f(p)
+ \ovpsi_1(-p) \hp_- \psi_1 (p) + \ovpsi_2 (-p) \hp_+ \psi_2 (p)},
\end{align}
where the dimensional lattice momentum is
$\hp_\pm = \frac{2}{a} \sin \frac{ap_\pm}{2}$.

\section{Ward-Takahashi identities }

The equivalence under the fields redefinition  leads to the
following identities:
\begin{align}
\braket{\mathcal{O}} = \invs{\mathcal{Z}} \int \mathcal{D} [\Phi] \mathcal{O} [\Phi] e^{i \mathcal{S} [\Phi]}
&= \invs{\mathcal{Z}} \int \mathcal{D} [\Phi'] \mathcal{O} [\Phi'] e^{i \mathcal{S} [\Phi']}, \notag \\
&= \invs{\mathcal{Z}} \int \mathcal{D} [\Phi] \maruk{ \mathcal{O} [\Phi] + \delta \mathcal{O} [\Phi] }
e^{i \mathcal{S} [\Phi] + i\delta \mathcal{S}}, \notag \\
&= \braket{\mathcal{O}} + \braket{ \delta \mathcal{O} [\Phi]} + \braket{\mathcal{O} [\Phi] \delta \mathcal{S} [\Phi]} + \cdots.
\end{align}
where we assume that the functional measure is not anomalous under
the symmetry.

If  the action is invariant under the transformation: $\delta
\mathcal{S} [\Phi]=0$, we obtain the following identity:
\begin{align}
\braket{ \delta \mathcal{O} [\Phi]} &= 0 \label{WTid}.
\end{align}
To find nontrivial relations between two point functions, we
examine possible combinations of operators for $\mathcal{O}$. For
example if we choose $\mathcal{O} = \phi \ovpsi_1$ and $\delta$ as
lattice SUSY transformation of $Q^{(+)}_+$, we obtain
\begin{align}
\braket{\psi_1(p) \ovpsi_1 (-p)} + \hp_+ \braket{\varphi (p) \ovphi (-p)} &= 0.
\label{WTid1}
\end{align}
Tree propagators are given by
\begin{align}
\braket{\varphi (p)  \ovphi (-p)}_{\text{tree}}= \frac{-1}{D(\hp)},  ~~~~
\braket{\psi_1 (p) \ovpsi_1 (-p) }_{\text{tree}} =
 \frac{\hp_+}{D(\hp)},
\label{2dprop}
\end{align}
where $D(\hp)=\hp_+\hp_- - m^2$.
Apparently tree propagators (\ref{2dprop}) satisfy the
identity (\ref{WTid1}), and it is consistent with the fact that the action
is exactly invariant under the lattice supersymmetry
at the classical level. We can choose other
combinations of fields and lattice super charges for examining the W-T identities.

The basic structure of the loop contribution of corresponding diagrams
to the two point function has the following form:
\begin{align*}
\braket{\varphi(p) \ovphi(-p)}_A &= \braket{\varphi(p) \ovphi(-p)}_{\text{tree}} X_A(\hp), & \\
\braket{\psi_1(p) \ovpsi_1(-p)}_A &= \braket{\psi_1(p) \ovpsi_1(-p)}_{\text{tree}} X_A(\hp),
\end{align*}
where $X_A(\hp)$'s are given as follows:
\begin{center}
\begin{tabular}{c||c|}
Loop diagram & ${ X_A(\hp)}$ \\ \hline \hline
\begin{minipage}{3cm}
\begin{tikzpicture}
\clip (-1.2,-0.1) rectangle (1.2, 1.1);
\draw[thick] (0, 0.5) circle (5mm);
\draw[thick] (-1,0) -- (1,0);
\node[left] at (-1,0) {};
\node[right] at (1,0) {};
\end{tikzpicture}
\end{minipage} & $0$ \\ \hline
\begin{minipage}{3cm} \begin{tikzpicture}
\clip (-1.2,-0.6) rectangle (1.2, 0.6);
\draw[thick] (0, 0) circle (5mm);
\draw[thick] (-1,0) -- (-.5, 0);
\draw[thick] (.5,0) -- (1, 0);
\node[left] at (-1,0) {};
\node[right] at (1,0) {};
\end{tikzpicture} \end{minipage}&
$\displaystyle -2g_3^2 \frac{\hp_+\hp_- + m^2}{D(\hp)} I_1 $  \\ \hline
\begin{minipage}{3cm} \begin{tikzpicture}
\clip (-1.2,-0.1) rectangle (1.2, 1.1);
\draw[thick] (0, 0.25) ellipse (4mm and 2.5mm);
\draw[thick] (0, 0.75) ellipse (4mm and 2.5mm);
\draw[thick] (-1,0) -- (1,0);
\node[left] at (-1,0) {};
\node[right] at (1,0) {};
\end{tikzpicture} \end{minipage} & $0$  \\ \hline
\begin{minipage}{3cm} \begin{tikzpicture}
\clip (-1.2,-0.6) rectangle (1.2, 0.6);
\draw[thick] (-1,0) -- (1,0);
\draw[thick] (0,0) circle (5mm);
\node [left] at (-1,0) {};
\node [right] at (1,0) {};
\end{tikzpicture} \end{minipage} &
$\displaystyle -6g_4^2 \frac{\hp_+\hp_- + m^2}{D(\hp)} I_2 $  \\ \hline
\begin{minipage}{3cm} \begin{tikzpicture}
\clip (-1.2,-0.6) rectangle (1.2, 0.6);
\draw[thick] (-1,0) -- (-0.5,0);
\draw[thick] (0.5, 0) -- (1,0);
\draw[thick] (0, 0.5) -- (0, -0.5);
\draw[thick] (0,0) circle (5mm);
\node [left] at (-1,0) {};
\node [right] at (1,0) {};
\end{tikzpicture} \end{minipage} &
$\displaystyle 16m^2 g_3^4 \frac{2 \hp_+ \hp_- + m^2}{D(\hp)} I_3 $  \\ \hline
\begin{minipage}{3cm} \begin{tikzpicture}
\clip (-1.2,-0.3) rectangle (1.2, 0.9);
\draw[thick] (-1,0) -- (1,0);
\draw[thick] (0.5, 0) arc (0:60:5mm);
\draw[thick] (-0.5,0) arc (180:120:5mm);
\draw[thick] (0,0.4) circle (2.5mm);
\node [left] at (-1,0) {};
\node [right] at (1,0) {};
\end{tikzpicture} \end{minipage}
& $\displaystyle 8g_3^4 \frac{\hp_+ \hp_- +m^2}{D(\hp)} I_4 $ \\ \hline
\end{tabular}
\end{center}
where
\begin{align}
I_1 &=\int \frac{d^2 \hk}{(2\pi)^2} \invs{D(\hk) D(\hp-\hk)},\\
I_2 &= \int \frac{d\hk_1^2}{(2\pi)^2} \frac{d\hk_2^2}{(2\pi)^2} \invs{D(\hk_1) D(\hk_2) D(\hp-\hk_1-\hk_2)}, \\
I_3 &= \int \frac{d\hk_1^2}{(2\pi)^2} \frac{d\hk_2^2}{(2\pi)^2} \invs{D(\hk_1) D(\hk_2) D(\hk_1+\hp) D(\hk_2+\hp) D(\hk_1- \hk_2)},\\
I_4 &= \int \frac{d^2\hk_1 d^2\hk_2}{(2\pi)^2(2\pi)^2 }
\frac{\hk_1^2+m^2}{D(\hk_1)^2 D(\hk_2)} \int \frac{d^2\hk}{(2\pi)^2}\invs{D(\hk) D(\hk_1-\hk)}
\end{align}

Therefore the W-T identity of this particular combination is
exactly satisfied up to the 2-loop level. We can show that the
other combinations of the two point functions and SUSY
transformations have the same structure as this example. In this
way we may conclude that the W-T identities are satisfied exactly
at the quantum level for all super charges. The details of the W-T
identities calculations for D=N=2 Wess-Zumino model will be found
in \cite{ADKK2012}.

\section{Discussions}

In confirming the exact lattice SUSY invariance lattice momentum
conservation plays a crucial role. This lattice momentum consevation
defines a new type of $\star$-product of fields $F$ and $G$:
\beq
 (F \star G)(p)
= \int d^2 p_1 d^2 p_2 F(p_1) G(p_2)
\delta^{(2)}(\hat{p}-\hat{p}_1-\hat{p}_2),   \label{starprod} \eeq
where the lattice momentum conservation is introduced. If we
introduce standard momentum $p$ conservation instead of the
lattice momentum $\hp$, the coordinate representation of the
product of the function $F$ and $G$ leads to the standard product.
However the coordinate representation of the $\star$-product with
the lattice momentum leads to a non-local product of two
functions. The details of $\star$-product can be found in
\cite{D'Adda:2010pg} and \cite{D'Adda:2012}. It would be
interesting to find a connection with this nonlocal nature of the
$\star$-product and the noncommutative nature of link approach of
lattice SUSY formulation\cite{DKKN} with Hopf algebraic lattice
SUSY invariance\cite{DKS}.

One of the other characteristics of this $\star$-product is that
the product is not associative. A given product, however, is well
defined and thus the invariance of the lattice SUSY transformation
is assured since SUSY transformation is linear with respect to
fields. However non-associativity may be a problem when we try to
extend this formulation to gauge theories since gauge
transformations are nonlinear in fields. We will come back to this
problem in future publication.
 Translational invariance is mildly broken since we use lattice
momentum which is not periodic in itself. We can, however, show that it
is recovered in the continuum limit\cite{D'Adda:2012}.

\end{document}